\pdfoutput=1
\documentclass[journal,twoside,web]{ieeecolor}
\usepackage{generic}
\usepackage{cite}
\usepackage{amsmath,amssymb,amsfonts}
\usepackage{algorithmic}
\usepackage{graphicx}
\usepackage{textcomp}
\def\BibTeX{{\rm B\kern-.05em{\sc i\kern-.025em b}\kern-.08em
    T\kern-.1667em\lower.7ex\hbox{E}\kern-.125emX}}
\begin{document}
\title{Nonvolatile Spintronic Memory Cells for Neural Networks}
\author{Andrew W. Stephan, Qiuwen Lou, Michael Niemier, X. Sharon Hu, \IEEEmembership{Fellow, IEEE} and Steven J. Koester, \IEEEmembership{Fellow, IEEE}.
\thanks{Manuscript submitted \today. This work was supported by Seagate Technology PLC.}
\thanks{The authors acknowledge the Minnesota Supercomputing Institute (MSI) at the University of Minnesota for providing resources that contributed to the research results reported within this paper. URL: http://www.msi.umn.edu}
\thanks{A. W. Stephan is with the College of Science and Engineering, University of Minnesota, Minneapolis, MN 55455 USA (e-mail:steph506@umn.edu).}
\thanks{Q. Lou is with the Department of Computer Science and Engineering, University of Notre Dame, Notre Dame, IN 46556 USA (e-mail:qlou@nd.edu).}
\thanks{M. Niemier is with the Department of Computer Science and Engineering, University of Notre Dame, Notre Dame, IN 46556 USA (e-mail:mniemier@nd.edu).}
\thanks{X. Sharon Hu is with the Department of Computer Science and Engineering, University of Notre Dame, Notre Dame, IN 46556 USA (e-mail:shu@nd.edu).}
\thanks{S. J. Koester is with the College of Science and Engineering, University of Minnesota, Minneapolis, MN 55455 USA (e-mail:skoester@umn.edu).}}

\maketitle

\begin{abstract} A new spintronic nonvolatile memory cell analogous to 1T DRAM with non-destructive read is proposed. The cells can be used as neural computing units. A dual-circuit neural network architecture is proposed to leverage these devices against the complex operations involved in convolutional networks. Simulations based on HSPICE and Matlab were performed to study the performance of this architecture when classifying images as well as the effect of varying the size and stability of the nanomagnets. The spintronic cells outperform a purely charge-based implementation of the same network, consuming $\approx$ 100 pJ total per image processed.
\end{abstract}

\begin{IEEEkeywords}Cellular Neural Network, Convolutional Neural Network, Spintronics, CMOS, Magnetoelectric, Rashba-Edelstein, MNIST, Nonvolatile Memory.
\end{IEEEkeywords}

\section{Introduction}
\label{sec:introduction}
A hardware implementation of feed-forward neural networks must incorporate three basic functionalities: a dot-product engine that can be used to compute convolution and fully-connected layer operations, memory elements that can be used to store intermediate inter- and intra-layer results, and components that can compute some non-linear activation function. Many purely charge-based implementations of these, with varying levels of efficiency, have been proposed. The dot-product has been successfully implemented in hardware in various ways\cite{OTAs,SAC,GDOT,Roy}. Using these dot-product circuits, a cellular neural network-based (CeNN) convolutional neural network (CoNN) accelerator was designed in\cite{CeNNs with CoNNs}. We propose a CeNN cell based on recently proposed spintronic elements with a high energy-efficiency that can be used as analog memory with a built-in activation function. The performance of these cells is simulated in a CeNN-accelerated CoNN performing image classification based on \cite{CeNNs with CoNNs}. The spintronic cells significantly reduce the energy and time consumption relative to their charge-based counterparts, needing only $\approx$ 100 pJ and $\approx$ 42 ns to compute all but the final fully-connected CoNN layer while maintaining a high accuracy.

\section{IRMEN}
\label{sec:IRMEN}
\subsection{IRMEN Neurons}
\begin{figure}
\centering
\includegraphics[scale=0.45]{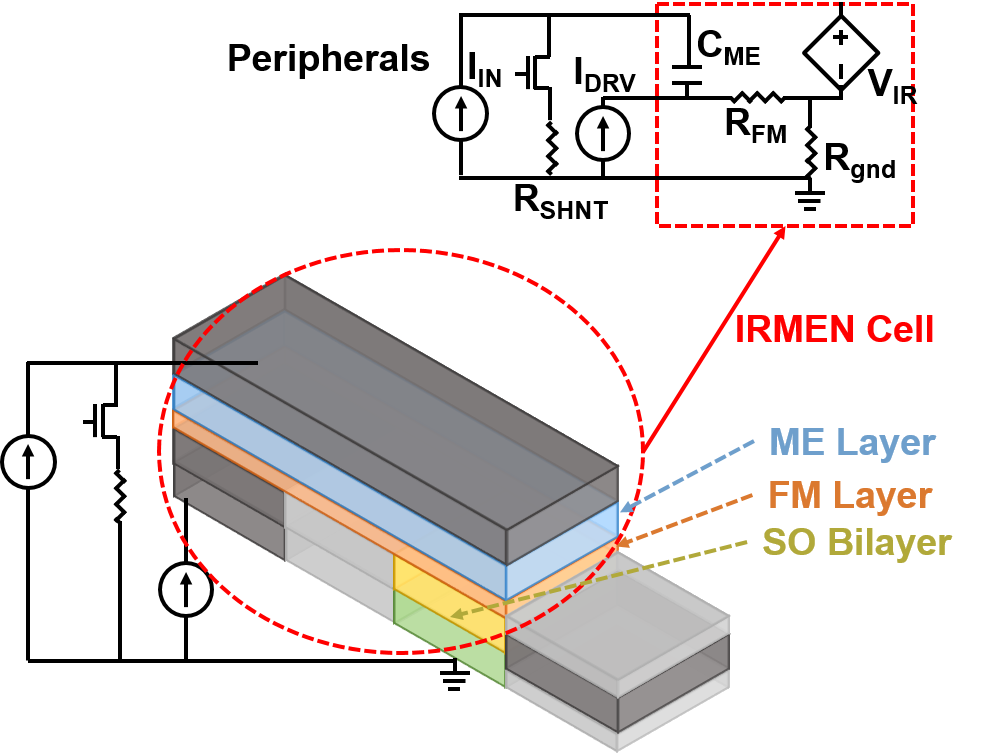}
\caption{Notional circuit diagram for an IRMEN-based CeNN cell utilizing the current-based input and shunt resistor model of the original cell proposed by Chua and Yang\cite{CNNs}. A 3D stack diagram of the spintronic portion of the cell is included.}
\label{fig:IRMENCell}
\end{figure}

\begin{figure}
\centering
\includegraphics[scale=1]{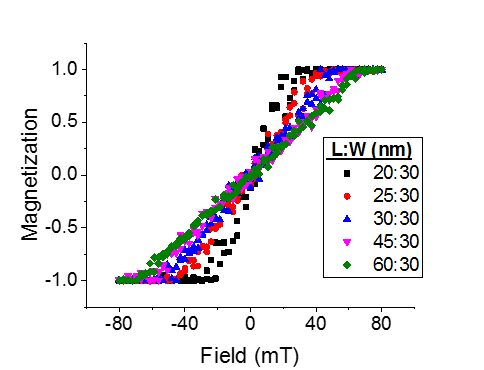}
\caption{The transfer function computed by the IRMEN cell on its inputs. The curve varies with the length/width ratio of the ferromagnet.}
\label{fig:Transfer}
\end{figure}

The Inverse Rashba-Edelstein Magnetoelectric Neuron (IRMEN) is thoroughly described, including its relationship to standard CeNN cells, in \cite{IRMEN}. A brief summary will be given here. The cell bears some resemblance to the standard cell of a cellular neural network in that it is based around a capacitor (see Fig. \ref{fig:IRMENCell})\cite{CNNs}. However the capacitor represents an input mechanism rather than the true state. A magnetoelectric material within the capacitor is coupled to the ferromagnet that makes up one of its electrodes, thereby allowing control of the ferromagnet via electric charge on the capacitor\cite{ME1,ME2,MESO1,MESO2,IRMEN}. Readout is accomplished by driving a charge current first through the ferromagnet to spin-polarize the current and then through an inverse spin-orbit stack that transduces from spin current to charge potential along an axis orthogonal to both current flow and spin orientation\cite{IR1,IR2,IR3}. This is modeled as a voltage source with the inverse Rashba potential $V_{IR}$ and internal resistance $R_{IR}$\cite{IRMEN}. The IRMEN cells natively compute a nonlinear transfer function on their inputs by virtue of the M-H curve dictated by the anisotropy of the nanomagnet. This transfer function can be tuned by varying the anisotropy characteristics. Example transfer functions are shown in Fig. \ref{fig:Transfer} assuming uniaxial crystalline anisotropy along the length dimension. 

\subsection{IRMEN Nonvolatile Memory}

\begin{figure}
\centering
\includegraphics[scale = 0.7]{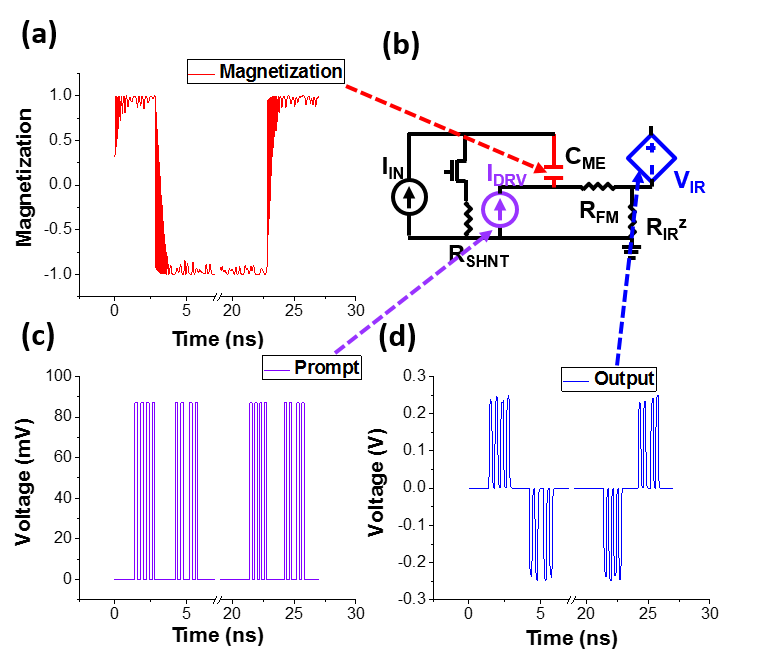}
\caption{The IRMEN memory cell in operation. (a) The magnetization is held at one of two different levels for an extended period. The device circuit model is shown in (b) for reference. (c) The device is prompted repeatedly to generate an output potential shown in (d) that gives a readout of the magnetization. This readout is slightly damped by the RC delay of the readout capacitance which partially smooths the thermal fluctuations of the magnet.}
\label{fig:MemoryPlots}
\end{figure}

\begin{figure}
\centering
\includegraphics[scale=0.5]{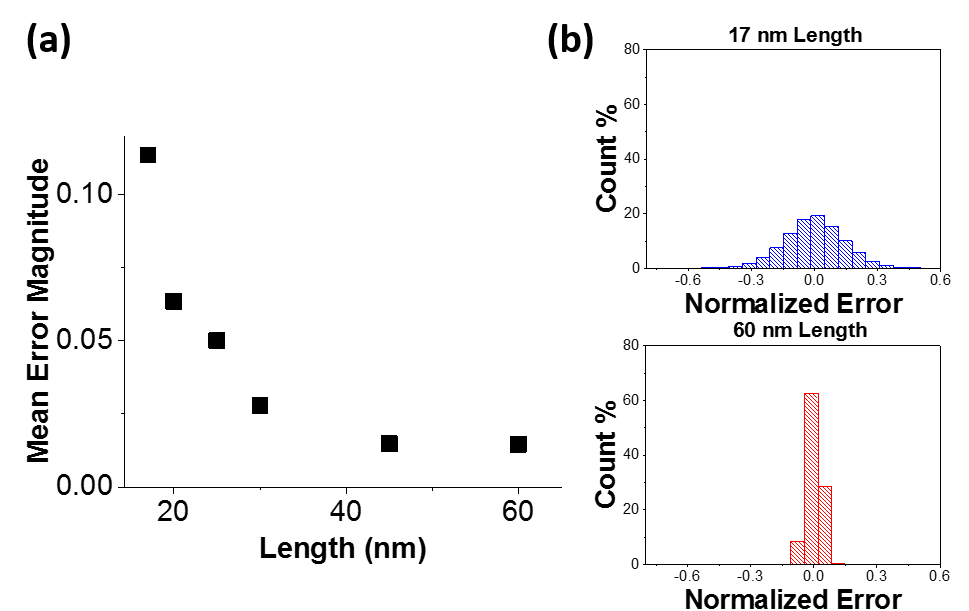}
\caption{(a) The mean magnitude of error vs. nanomagnet length. (b) Histogram of errors for lengths of 17nm and 60 nm, respectively. The increased range of errors in the former is due to higher levels of thermal noise.}
\label{fig:ErrorHist}
\end{figure}

Memory is a crucial component to any hardware-based implementation of a neuromorphic computing scheme. In anything more complex than a simple fully-connected network, such as a CoNN, the input values to any given layer need to be referenced multiple times\cite{CoNNs}. This necessitates the inclusion of some form of analog memory or digital memory accompanied by analog-to-digital converters (ADCs) and the reverse (DACs). The simplest solution is to use simple 1T DRAM cells. However one of the issues from which standard 1T DRAM suffers is the loss of charge upon read. If the state is not refreshed the leakage of repeated read cycles will degrade the stored state. The device discussed above, besides being able to realize neuromorphic CeNN-like operations, can also function as a memory. Because it serves dual purpose as a neural computing unit and analog memory, the IRMEN offers enhanced efficiency. Here the state is stored on the capacitor. The IRMEN readout mechanism, crucially, does not involve accessing the charge stored on the cell capacitor. This charge may remain safely locked in place while the cell state is read out through injection of current into the ferromagnet (FM). While this does temporarily disturb the actual electrical field across the capacitor and therefore the actual steady-state magnetization, a sufficiently swift read pulse will ensure the read completes while the FM is just beginning to move depending on the relative time-scales of electrical and magnetic motion. More importantly, the cell will return to the same state after any perturbations caused by the read without any loss of information. Subsequent reads will obtain the same actual value as the first. The perturbation during read, even if non-negligible, will be consistent and can be accounted for. We note that the intrinsic write energy of the IRMEN memory is equal to the charging energy of the magnetoelectric capacitor. Using values from Table \ref{tab:parameters}, we estimate the write energy to be no more than 10 aJ, putting it on par with typical 1T DRAM values. This value scales with the capacitor area. This estimate does not include the energy dissipated by external circuitry during the write process. Again referencing Table \ref{tab:parameters} we estimate the total write energy at between 0.24 and 0.79 fJ for one cell. While this is higher than some other nonvolatile memories such as STT-RAM, this is compensated for by the analog, as opposed to digital, nature of the IRMEN cell as well as its ability to function as a neural computing unit in addition to memory\cite{STT-RAM}. 

A basic simulated demonstration of the functionality of the IRMEN memory cell is shown in Fig. \ref{fig:MemoryPlots}. The magnetization is switched between two different levels, held for an extended period of 12 ns by charge stored on the capacitor, and switched again. The drive current is pulsed for periods of 200 ps to generate a readout potential at various points throughout the simulation. This indicates the ability of the IRMEN to store a value, encoded as charge, and provide a transformed readout value on command. We note that there is some noise in the readouts visible in Fig. \ref{fig:MemoryPlots} (d). In order to quantify this error we performed more exhaustive simulations. For nanomagnets of several different sizes, Monte Carlo simulations of 12300 iterations were performed. In each iteration an input current was applied and the neuron was then prompted to provide a readout. The results approximated a saturated linear transfer function as in the ideal CeNN cell\cite{CNNs} with some error. After normalizing according to the base input current and output voltage values as described in Sec. \ref{sec:Simulation}, the error is calculated as the difference between an iealization of the transfer function and the actual readout. The average magnitude of the normalized error is shown in Fig. \ref{fig:ErrorHist} (a). Histograms indicating the range of errors for the largest and smallest nanomagnets are shown in Fig. \ref{fig:ErrorHist} (b), displaying the significant increase in thermal broadening with smaller and more unstable magnets. 

The following section will describe a network of IRMEN cells for CoNN implementation.

\section{Convolutional Networks with IRMEN Cells}
\label{sec:CeNNs with CoNNs}
\begin{figure*}
\centering
\includegraphics[scale=.73]{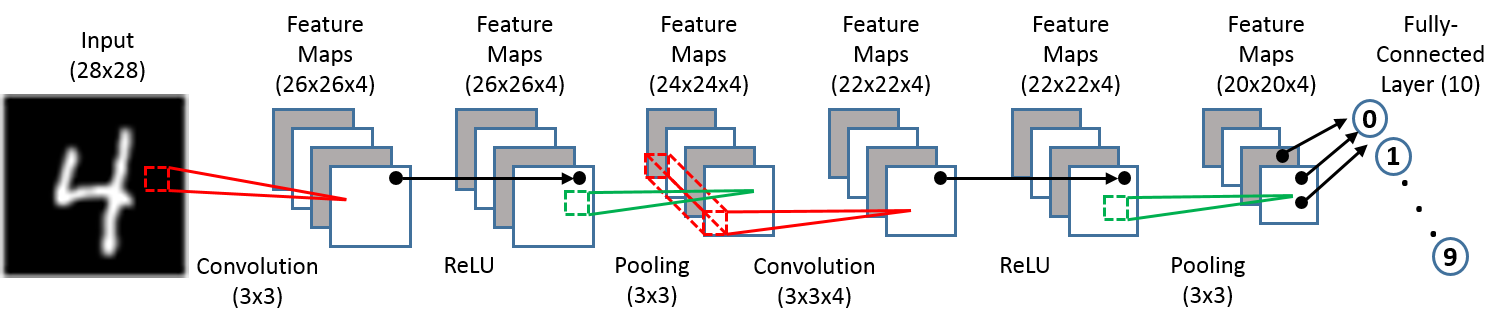}
\caption{Convolutional neural network structure used to classify handwritten digits. Tensorflow was used to train weights before converting to the CeNN-based CoNN accelerator as described in \cite{CeNNs with CoNNs}.}
\label{fig:CoNNStructure}
\end{figure*}

Lou \emph{et al.} proposed using standard, purely charge-based CeNN cells with weighted, programmable Operational Transconductance Amplifier-based current sources (OTAs) to produce an in-hardware implementation of a convolutional neural network \cite{CeNNs with CoNNs,OTAs}. Using CeNN weight template schemes, the Rectified Linear Unit (ReLU) activation function and pooling function were approximated so that all but the fully-connected output layer of a CoNN could be implemented via simple CeNN cells. We propose to improve on the purely charge-based CeNN/CoNN model by replacing the purely charge-based cells with a more energy-efficient IRMEN cell. The CoNN structure to be emulated is shown in Fig. \ref{fig:CoNNStructure}.

\subsection{Coupled CeNNs for Memory and Computing}
\begin{figure}
\centering
\includegraphics[scale=0.45]{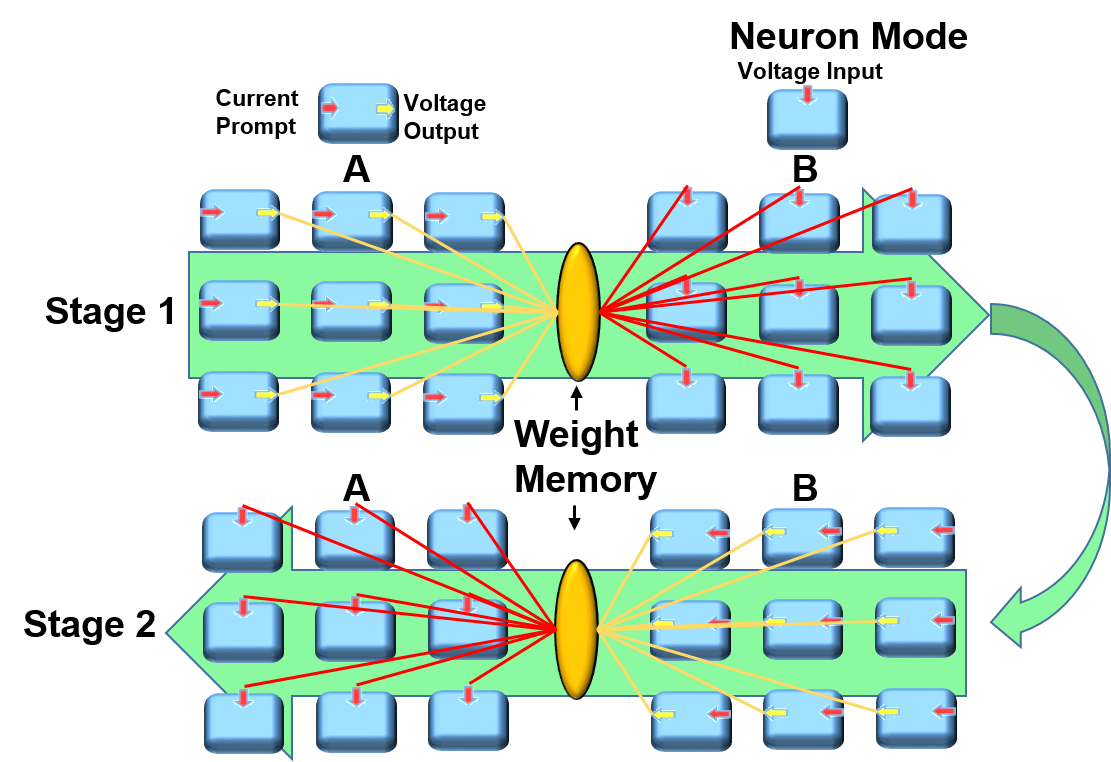}
\caption{Diagram of the dual-network system implemented to leverage the IRMEN memory. The IRMENs are represented here as objects that read in a voltage, remember it, and provide an output voltage as a function of the input when prompted by an injected current. Two stages of computation are shown with the background arrow indicating the direction of information flow. In stage 1 network A acts as memory cells for B; in stage 2 the A acts as memory for B.}
\label{fig:DualNetwork}
\end{figure}

To fully leverage the IRMEN capabilities we propose a dual-functionality operation mode. Each stage corresponds to one CeNN operation, represented by a specific CeNN template imposed upon the data as it is slowly transformed from the initial input to the final output. Each stage of computation is implemented by one of a pair of identical IRMEN CeNNs similar to the structure in \cite{CeNNs with CoNNs}. One of the pair provides the input, obtained from the previous stage and locally stored, to the other of the pair for processing and subsequent storing (see Fig. \ref{fig:DualNetwork}).

The proposed dual-CeNN design allows us to do away with a dedicated memory module for neuron state storage and most of the associated peripheral circuits such as ADCs and DACs. Weight storage must still be implemented, although additional IRMEN memory cells could be included for this purpose. We expect significant savings in operational energy and possibly also speed as a result. Although the delay between stages must be on the order of nanoseconds to allow sufficient time for the magnetizations to respond to the charge placed on the capacitors, each stage only needs to be powered for about 130 ps, which is the estimated combined delay of the OTAs\cite{OTAs} and other electrical components. The total operational delay between each stage is 1.5 ns, which is mostly due to the magnetic switching delay as previously mentioned.

\section{Simulation}
\label{sec:Simulation}
This section will describe the methods by which we simulate IRMEN CeNNs emulating an image classification CoNN as previously described. The results of this simulation as measured by the predicted accuracy and energy cost will also be presented herein.

\begin{table}
\centering
\caption{Simulation Parameters}
\label{table}
\setlength{\tabcolsep}{3pt}
\begin{tabular}{|p{30pt}|p{80pt}|p{60pt}|}
\hline 
Symbol&
Quantity& 
Value \\
\hline
\vspace{0.005in}
$K$&
\vspace{0.005in}
crystalline anisotropy& 
\vspace{0.005in}
$4.5 \cdot 10^4$ J/m$^3$ \\
$V$ &
ferromagnet volume& 
1530 - 5400 nm$^3$ \\
$\Delta$ &
thermal stability factor&
16 - 59 kT \\
$\eta$&
spin injection efficiency& 
0.8\cite{Heusler1, Heusler2, Heusler3} \\\
$M_S$&
saturated magnetization& 
1.7 MA/m \\
$\alpha$&
Gilbert damping &
0.01 \\
$C$ &
ME capacitance & 
1 fF \\
$\alpha_{ME}$ &
ME coefficient &
10/\textit{c}*\cite{ME2, HighME1} \\
$\lambda$&
spin conversion length& 
6 nm\cite{HighIR1,HighIR2,HighIR3} \\
$\rho$&
IR material resistivity& 
10 m$\Omega$ $\cdot$ cm\cite{HighResistivity} \\
$R_{IR}$ &
IR source resistance &
20 k$\Omega$ \cite{IRMEN}\\
$V_D$&
neuron drive voltage& 
$\le$ 100 mV \\
$V_S$&
synapse supply voltage & 
$\pm$ 500 mV \\
$V_T$&
transistor threshold &
0.2 V\\
\hline
\multicolumn{3}{p{200pt}}{*\textit{c} is the speed of light.}\\
\end{tabular}
\label{tab:parameters}
\end{table}

\subsection{Simulation Setup}
In order to evaluate the utility of the proposed IRMEN-based architecture we tested it on the MNIST handwritten digits dataset. The structure of the CoNN to be implemented via IRMEN CeNNs is shown in Fig. \ref{fig:CoNNStructure}, is identical to one of the test cases in \cite{CeNNs with CoNNs}. In the interest of time the training of the weights was implemented using TensorFlow with a custom transfer function representing a close numerical approximation to the hysteresis curve of the IRMENs' nanomagets. The trained weights were then given to a Matlab simulator which used the fourth-order Runge-Kutta method to simultaneously solve the Landau-Lifshitz-Gilbert equation and the electrical circuit equations associated with each neuron. The IRMEN magnetodynamical equations used were identical to those of \cite{IRMEN}. The FM is modeled using the macrospin approximation. The Landau-Lifshitz-Gilbert equation relates the motion of the unit magnetization $\boldsymbol{m}$ to the net field
\begin{gather}
\frac{d\boldsymbol{m}}{dt} = -\gamma \mu_0 \Big( (\boldsymbol{m} \times \boldsymbol{H_{Eff}}) - \alpha \big( \boldsymbol{m} \times (\boldsymbol{m} \times \boldsymbol{H_{Eff}}) \big) \Big),
\end{gather}
where $\gamma$ is the gyromagnetic ratio, $\mu_0$ is the vacuum permeability, $\alpha$ is the Gilbert damping and $\boldsymbol{H_{Eff}}$ is the effective field, calculated according to the methods in \cite{IRMEN}. The simulation parameters are given in Table \ref{tab:parameters}.

In order to map from the Tensorflow network to an IRMEN simulation we must translate the arbitrary state and weight values into electrical values associated with the IRMEN cells. A neuron state value can vary between $\pm$ 1, depending on the magnetization of the associated nanomagnet, and is read by measuring $V_{IR}$. This value depends on the driving potential in addition to the state of the nanomagnet. We define the maximum output potential $V_1$ to be the neuron state of 1:
\begin{gather}
V_1 = V_D \frac{R_{IR}}{R_{FM} + R_{gnd}} \frac{\eta}{\lambda}.
\end{gather}
\label{V1}
The normalized output of a neuron is $V_{IR}/V_1$. This voltage is applied to the input of an OTA which consequently produces up to 1 $\mu$A of current to inject through the shunt resistor of a subsequent cell (see Fig. \ref{fig:IRMENCell}). The normalized input to a neuron is thus $I_{IN}/(10^{-6})$. We use a similar two-stage OTA design as \cite{CeNNs with CoNNs}. The first stage in the OTA is a differential input pair. The second stage is a current mirror that subtracts two branches of current to obtain a single ended output while providing a large output impedance. The ratio of the current mirrors between two stages is set to 2 to save power in the first stage of the OTA. We use multiple OTAs to represent different numbers of bits for weights by power gating. By reprogramming these OTAs, the desired weights are achieved in each step.

\subsection{Image Classification Results}

\begin{figure}
\centering
\includegraphics[scale=0.45]{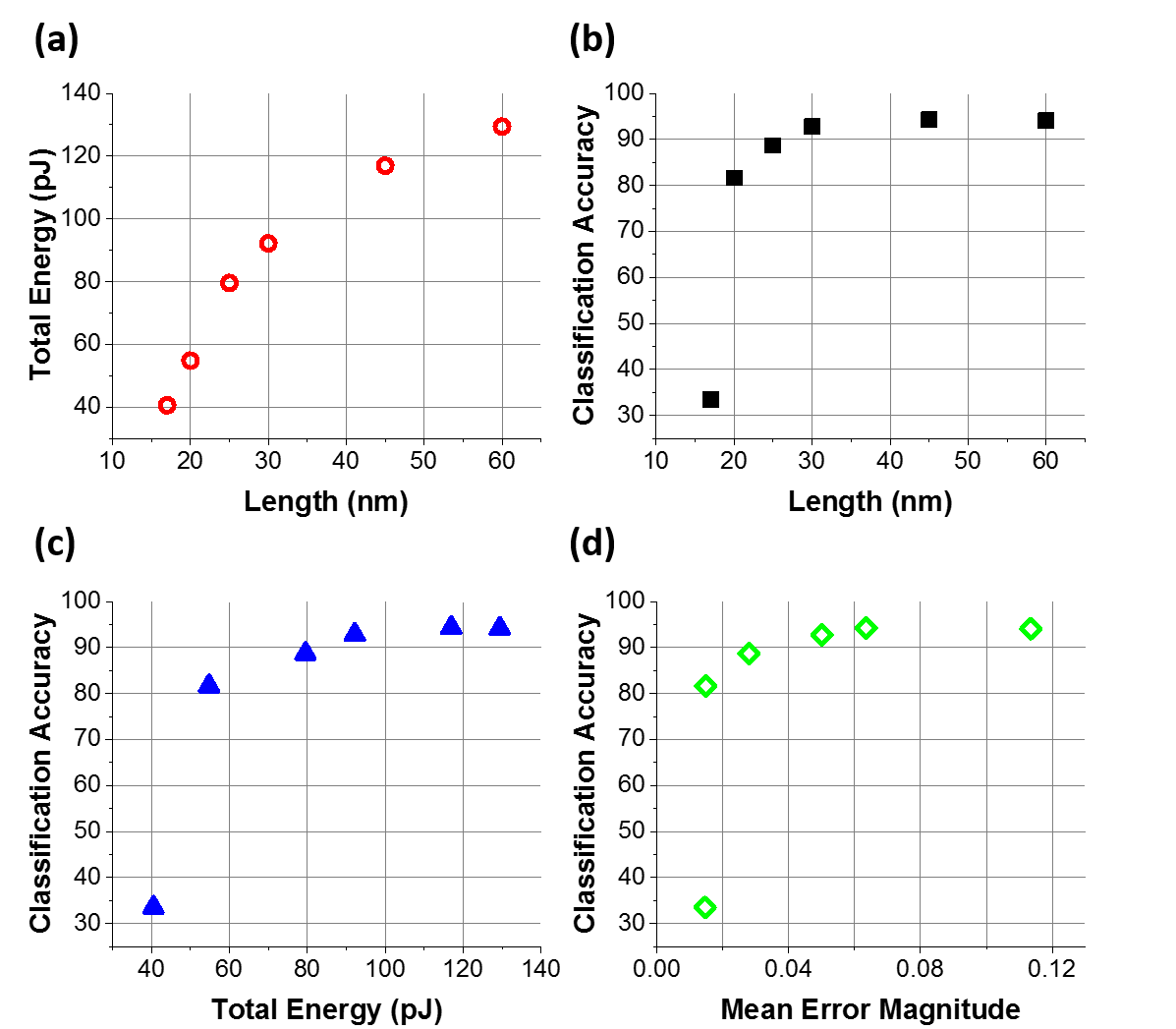}
\caption{IRMEN CeNN/CoNN image classification results. (a) Total energy consumed (excepting fully-connected layer) vs. the length of the cell FMs. (b) Image classification accuracy vs. FM length. (c) Classification accuracy vs. total energy. (d) Classification accuracy vs. the mean magnitude of the normalized transfer function accuracy computed by the FMs in the network, as influenced by size and thermal noise.}
\label{fig:ED}
\end{figure}

\begin{figure}
\centering
\includegraphics[scale=0.6]{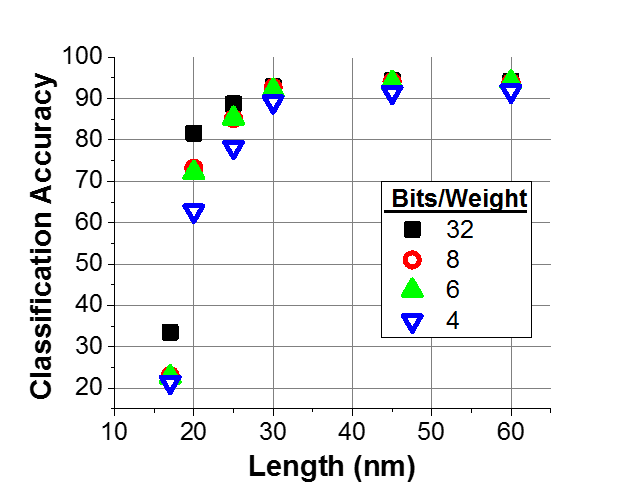}
\caption{IRMEN CeNN/CoNN image classification results plotted vs. the length of the cell FMs with different levels of weight representation precision starting from 4 bits up to essentially perfect precision.}
\label{fig:Bits}
\end{figure}

The simulated CoNN with IRMEN-based convolution, activation and max-pooling templates was tested against 10000 images withheld from the training stage and achieved high image classification accuracy (see Fig. \ref{fig:ED}). The energy required to perform all the steps involved in the classification of each image, except the final fully-connected layer, was estimated, including cycling the OTA and weight storage access transistors, cycling the OTA gate capacitors, powering the OTAs themselves and driving the current through the IRMEN cells. This total varied depending on IRMEN FM length, as greater lengths require larger magnetoelectric fields to saturate the FM along the competing axis (see Fig. \ref{fig:Transfer}). The relationship between energy and length is shown in Fig. \ref{fig:ED} (a). The classification accuracy is plotted against FM length in Fig. \ref{fig:ED} (b) showing that greater thermal stability improves overall network performance. However this comes at the cost increased energy and area. The correlation between accuracy and energy expenditure is displayed in Fig. \ref{fig:ED} (c). The origin of this increased accuracy is the reduced error in the saturated linear transfer function computed by each IRMEN cell (see Fig. \ref{fig:Transfer}). The classification accuracy is plotted directly against the mean transfer function error magnitude in Fig. \ref{fig:ED} (d). We also consider the effect of using weights with limited resolution. Although the specific weight storage mechanism is not considered here, it is reasonable to assume a limited representation accuracy. Thus in Fig. \ref{fig:Bits} we show the image classification accuracy vs. FM length with the number of bits used to encode a weight value as a parameter, starting with 4-bit precision. Although lowering the precision to 4 bits is detrimental the performance is still quite good for lengths above 30 nm. 

Compared to the previously-implemented purely charge-based version, using the IRMEN cells provides a significant energy and time savings. According to Table 3 of \cite{CeNNs with CoNNs} the charge-based CeNN requires over 12 nJ to compute all convolution, pooling and activation stages, while the IRMEN CeNN needs less than 0.14 nJ. We also note that the Convolution/ReLU and Pooling layers require two and twelve individual CeNN stages, respectively, each with a delay of 1.5 ns (see Sec. \ref{sec:CeNNs with CoNNs} A.) so the overall IRMEN CeNN/CoNN delay is 42 ns. The CeNN in \cite{CeNNs with CoNNs} takes 240.5 ns to perform the same computations. 

\section{Conclusion}
\label{sec:Conclusion}
With the growing importance of neuromorphic computing and beyond-CMOS computation, the search for new devices to fill these roles in crucial. We have proposed a novel magnetoelectric analog memory element with a built-in transfer function that also allows it to act as the cell in a CeNN. Simulations of a CeNN-friendly CoNN implemented with these IRMEN cells predict that highly accurate and low-power image classification can be achieved with this device. These results are superior to those predicted for purely charge-based implementations of the same network. This clearly demonstrates the benefits of applying spintronics to neurmorphic computing.

\end{document}